\documentclass[aps,english,twocolumn]{revtex4}

\usepackage{graphicx}
\usepackage{amsmath, amssymb}
\usepackage{babel}

\begin{document}

\title{
Doping dependent charge transfer gap and realistic electronic model
of n-type cuprate superconductors }

\author{T. Xiang$^{1,2}$, H. G. Luo$^2$, D. H. Lu$^3$, K. M. Shen$^4$, Z. X. Shen$^3$}

\address{$^1$Institute of Physics, Chinese Academy of
Sciences, P.O. Box 603, Beijing 100080, China}

\address{$^2$Institute of Theoretical Physics, Chinese Academy of
Sciences, P.O. Box 2735, Beijing 100080, China}

\address{$^3$Department of Physics, Applied Physics,
and Stanford Synchrotron Radiation Laboratory, Stanford University,
Stanford, California 94305, USA}

\address{$^4$Department of Physics and Astronomy, University of British
Columbia, Vancouver, British Columbia, V6T 1Z4, Canada }

\date{\today}

\begin{abstract}
Based on the analysis of the measurement data of angle-resolved
photoemission spectroscopy (ARPES) and optics, we show that the
charge transfer gap is significantly smaller than the optical one
and is reduced by doping in electron doped cuprate superconductors.
This leads to a strong charge fluctuation between the Zhang-Rice
singlet and the upper Hubbard bands. The basic model for describing
this system is a hybridized two-band $t$-$J$ model. In the symmetric
limit where the corresponding intra- and inter-band hopping
integrals are equal to each other, this two-band model is equivalent
to the Hubbard model with an antiferromagnetic exchange interaction
(i.e. the  $t$-$U$-$J$ model). The mean-field result of the
$t$-$U$-$J$ model gives a good account for the doping evolution of
the Fermi surface and the staggered magnetization.
\end{abstract}

\maketitle

The evolution of the Fermi surface and the Mott insulating gap with
hole or electron doping is a central issue in elucidating the
mechanism of high-T$_c$ superconductivity. In the hole doped case, a
small Fermi surface arc appears first near $(\pi/2 ,\pi /2)$ and
then extends towards $(\pi , 0)$ and $(0, \pi )$ with increasing
doping. In contrast, in the electron dope case, electrons are first
doped into the upper Hubbard band (Cu $ 3d^{10} $ band) near ($\pi $
,0) and equivalent points. With further doping but still in the
antiferromagnetic phase, in-gap spectral weight develops below the
Fermi level. These in-gap states move upwards and eventually form a
hole-like Fermi surface pocket around ($\pi /2$ , $\pi
/2$)\cite{Armitage02}. In the heavily overdoped sample, these two
Fermi pockets merge together and form a large Fermi surface with a
volume satisfying the Luttinger theorem.

The peculiar doping dependence of the Fermi surface topology in
electron-doped cuprates is a manifestation of correlation effects.
To understand the physics behind, much of the theoretical study has
been carried out with the one-band Hubbard
model\cite{Kusko02,Kusunose03,Senechal04}. In this model, a metallic
band is split into two effective bands, namely the upper and lower
Hubbard bands, by a correlation energy $U$ that represents the
energy cost for a site to be doubly occupied. Under the mean-field
approximation, this model gives a good account for the experimental
data if $U$ is assumed to fall strongly with doping. However, this
strong reduction of $U$ by doping is not usually
expected\cite{Yuan04}. The nominal Hubbard $U$-term could arise
either from the on-site Coulomb repulsion between two electrons in a
Cu $3d_{x^2-y^2}$ orbital or from the charge transfer (CT) gap
between O $2p$ and Cu $3d^{10}$ bands. The on-site Coulomb repulsion
in a Cu $3d_{x^2-y^2}$ states in high-T$_c$ cuprates is generally
larger than $5$ eV. The CT gap in electron doped cuprates quoted
from the optical data is also quite large ($\sim 1.5$
eV)\cite{Onose01,Wang06}. It seems that in neither case $U$ can be
dramatically suppressed by only 15\% doping.

An alternative interpretation to the two Fermi pockets is based on
the notion of band folding induced by the antiferromagnetic
interaction\cite{Yuan04,Matsui05}. This interpretation is consistent
with the measurement data in the overdoped regime ($x > 0.15$).
However, in the low doping antiferromagnetic phase, it breaks down.
The band folding assumes implicitly a band with large Fermi surface
exists and it is the antiferromagnetic interactions between the hot
spots split this band into a conduction electron and a shadow hole
band. However, in the antiferromagnetic phase at low doping, these
bands with the folding gap at the hot spots were not observed and
the band near $(\pi /2, \pi /2)$ is well below $(\pi
,0)$.\cite{Armitage02} Furthermore, the antiferromagnetic
interaction is too small to account for the energy splitting between
the lower and upper CT bands at least in the low doping limit.

To resolve the above problems, it is important to understand
correctly how the hole-like Fermi pockets develop with doping. In a
nominally undoped Nd$_2$CuO$_4$, a dispersive band structure is
observed by ARPES at roughly $1.2$ eV below the chemical potential.
As shown in Ref. \cite{Armitage02}, the energy-momentum dispersion
of this spectral peak behaves almost the same as the lower CT band
observed in Ca$_2$CuO$_2$Cl$_2$, except in the latter case the band
lies at only $\sim 0.7$ eV below the chemical potential. This
suggests that these two bands have the same physical origin. The
difference is probably due to the intrinsic doping and the Fermi
energy is pinned near the bottom of conduction band (i.e. Cu
$3d^{10}$ band) in Nd$_2$CuO$_4$ and near the top of the valence
band (\emph{i.e.} Zhang-Rice singlet band\cite{Zhang88}) in
Ca$_2$CuO$_2$Cl$_2$.

\begin{figure}[ht]
\includegraphics[width=\columnwidth]{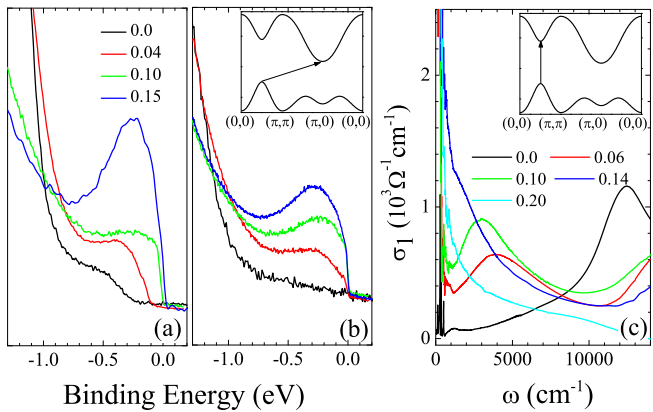}
\caption{ARPES spectra near (a) the nodal and (b) antinodal regions,
reproduced from the data published in Ref.~\cite{Armitage02}. (c)
The infrared conductivity reproduced from the data published in Ref.
\cite{Wang06}. The insets of (b) and (c) illustrate the indirect and
direct CT gaps. } \label{fig:infra}
\end{figure}

Doping electrons into Nd$_2$CuO$_4$ results in a spectral weight
transfer from the main spectral peak at $\sim 1.2$eV to a ``in-gap"
state. This in-gap state first appears as a week low energy ``foot"
at $\sim 0.5$ eV below the Fermi level $\varepsilon_F$ along the
zone diagonal in the undoped Nd$_2$CuO$_4$ (Fig.~\ref{fig:infra}).
It moves towards the Fermi level with doping and becomes a broad
hump just below the Fermi level at optimal doping. The hole Fermi
pockets observed at high doping originate from these in-gap states.
In contrast, the states near $(\pi,0)$ resides at $\varepsilon_F$ as
they are derived from the bottom of the upper Hubbard band. The fact
that the broad maximum is slightly below $\varepsilon_F$ is caused
by the Frank-Condon broadening as discussed below.

It should be pointed out that, same as for the dispersive high
energy band, the in-gap states behave similarly as the low energy
coherent states observed in hole-doped
Ca$_2$CuO$_2$Cl$_2$\cite{Shen04}. Near half filling, the in-gap
state in Nd$_2$CuO$_4$ lies also at $\sim 0.7$ eV above the high
energy spectral peak. This suggests that, similar as in hole doped
materials, the high energy hump structure in the spectra results
from the Franck-Condon broadening and the in-gap states are the true
quasiparticle excitations located at the top of the lower CT
band\cite{Shen04}. At half filling, the in-gap state is not observed
because its quasiparticle weight is vanishingly small\cite{Rosch05}.

The spectral weight transfer induced by doping has also been
observed in the optical measurements (Fig.~\ref{fig:infra}). At zero
doping, the optical CT gap appears at $\sim 1.5$ eV. Upon doping, a
mid-infrared conductivity peak develops. This mid-infrared peak
appears at $\sim 0.5$ eV at low doping\cite{Onose01,Wang06}, and
then moves towards zero energy with increasing doping. The doping
dependence of the mid-infrared peak is consistent with the doping
evolution of the ``in-gap'' states observed by ARPES. It suggests
that the mid-infrared peak results mainly from the optical
transition between the ``in-gap'' states and the upper Hubbard band.
The polaron effect may also have some contribution to these
mid-infrared peaks.\cite{Mishchenko08}

The above discussion indicates that the true CT gap, measured as the
minimum excitation energy between the lower and upper Hubbard bands,
is only $0.5$ eV at half filling, much lower than the optically
measured CT gap, which is usually believed to be about 1.5 eV. This
difference between the true quasiparticle gap that determines the
transport and thermodynamics and optically measured CT gaps has also
been found in hole doped materials. For La$_2$CuO$_4$, Ono \emph{et
al.}\cite{Ono07} found recently that the CT gap obtained from the
high temperature behavior of the Hall coefficient is only 0.89 eV,
while the corresponding optical CT gap is about 2 eV. This means
that the optical CT gap, which is generally determined from the peak
energy of the optical absorption, does not correspond to the true
gap between the two bands in high-T$_c$ oxides. The indirect nature
of the gap (insets of Fig.~\ref{fig:infra}) and the Frank-Condon
effect lead to the overestimate of the gap by optics. It also means
that the charge fluctuation in high-T$_c$ materials is much stronger
than usually believed and should be fully considered in the
construction of the basic model of high-T$_c$ superconductivity
\cite{Zaanen85, Varma06, Luo05}.

The doping dependence of low energy peaks observed by both APRES and
optics indicates that there is a gap closing with doping. This gap
closing may result from the Coulomb repulsion between O $2p$ and Cu
$3d$ electrons. Doping electrons will increase the occupation number
of Cu $3d$ states, which in turn will add an effective potential to
the O $2p$ states and raise their energy level. If $U_{pd}$ is the
energy of the Coulomb interaction between neighboring O and Cu ions,
then the change in the O $2p$ energy level will be $\delta
\varepsilon_{p} \approx + 2 x U_{pd} $, where $x$ is the doping
concentration and the factor 2 appears since each O has two Cu
neighbors. $U_{pd}$ is generally estimated to be of order 1-2 eV.
Thus a 15\% doping of electrons would reduce the CT gap by 0.3-0.6
eV, within the range of experimentally observed gap reduction.
Furthermore, the electrostatic screening induced by doping can
reduce the on-site Coulomb interaction of Cu $3d_{x^2-y^2}$
electrons. This can also reduce the gap between the O $2p$ and the
upper Hubbard bands.

\begin{figure}[ht]
\includegraphics[width=0.8\columnwidth]{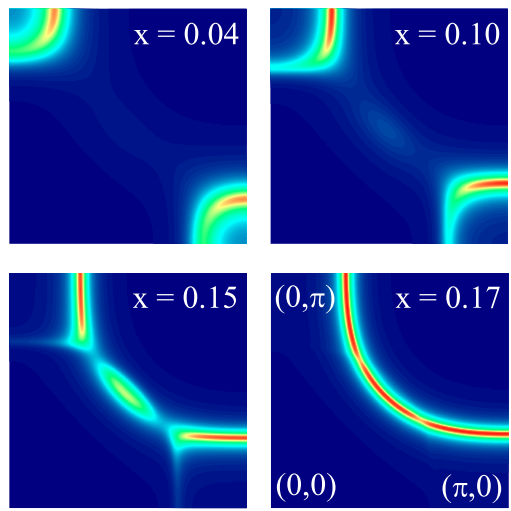}
\caption{Fermi surface density map at different dopings $x$,
obtained by integrating the spectral function from -40 to 20 meV
around the Fermi level, for the $t$-$U$-$J$ model. }
\label{fig:band}
\end{figure}

Now let us consider how to characterize the low energy charge and
spin dynamics of the system. For simplicity, we focus on the
electronic structure and leave the additional electron-phonon
interaction effect for future study. The lower CT band behaves
similarly as the Zhang-Rice singlet band. Thus if its charge
fluctuation with the upper Hubbard band is ignored, this band should
be described by an effective one-band $t$-$J$ model\cite{Zhang88}.
Similarly, the upper Hubbard band should also be described by an
effective one-band $t$-$J$ model if there is no charge fluctuation.
However, in the case the hybridization or charge transfer between
these bands is important, it can be shown from a three-band model
that these two $t$-$J$ models should be combined together and
replaced by the following hybridized two-band $t$-$J$
model\cite{Baskara}
\begin{eqnarray}
H &=&\sum_{ij\sigma }t_{ij}^{e}e_{i}^{\dagger }d_{i\sigma }d_{j\sigma
}^{\dagger }e_{j}+\sum_{ij\sigma }t_{ij}^{h}h_{i}^{\dagger }d_{i\sigma
}d_{j\sigma }^{\dagger }h_{j}  \notag \\
&&+\sum_{ij\sigma }t_{ij}\left( \sigma d_{i\sigma }^{\dagger }d_{j\overline{
\sigma }}^{\dagger }e_{i}h_{j}+h.c.\right) +J\sum_{\langle ij\rangle
}\mathbf{S}_{i}\cdot \mathbf{S}_{j}  \notag \\
&&+\sum_{i}\left( \varepsilon _{e}e_{i}^{\dagger }e_{i}+\varepsilon
_{h}h_{i}^{\dagger }h_{i}\right) -V_{pd}\sum_{\langle ij\rangle
}e_{i}^{\dagger }e_{i}h_{j}^{\dagger }h_{j},  \label{eq:tj}
\end{eqnarray}
where $h_{i}$, $e_{i}$ and $d_{i\sigma }$ are the annihilation
operators of a Zhang-Rice singlet hole, a doubly occupied
$d_{x^{2}-y^{2}}$ state (doublon), and a pure Cu$^{2+}$ spin,
respectively. At each site, these three states cannot coexist and
the corresponding number operators should satisfy the constraint
\begin{equation}
e_{i}^{\dagger }e_{i}+h_{i}^{\dagger }h_{i}+\sum_{\sigma }d_{i\sigma
}^{\dagger }d_{i\sigma }=1.
\end{equation}
The difference between the number of doubly occupied
$d_{x^{2}-y^{2}}$ states and Zhang-Rice singlet holes is the doping
concentration of electrons, $\langle e_{i}^{\dagger
}e_{i}-h_{i}^{\dagger }h_{i} \rangle =x$.

In Eq. (\ref{eq:tj}), $\mathbf{S}_{i}= d^\dagger_i \mathbf{\sigma}
d_i/2$ is the spin operator and $\mathbf{\sigma}$ is the Pauli
matrix. $\varepsilon _{e}$ and $\varepsilon_{h}$ are the excitation
energies of a doublon and a Zhang-Rice singlet, respectively.
$t_{ij}^{e}$ and $t_{ij}^{h}$ are the hopping integrals of the upper
Hubbard and Zhang-Rice singlet bands. In Eq. (\ref{eq:tj}), if
$\varepsilon_{h}\gg \varepsilon_{e}
> 0$, then $\langle h_{i}^{\dagger }h_{i} \rangle
\approx 0$ and $H$ simply reduces to the one-band $t$-$J$ model of
doubly occupied electrons in the doublon-spinon representation. On
the other hand, if $\varepsilon _{e}\gg \varepsilon_{h}>0$ , then
$\langle e_{i}^{\dagger }e_{i} \rangle \approx 0$ and $H$ becomes
simply the one-band $t$-$J$ model of Zhang-Rice singlets in the
holon-spinon representation. The $t_{ij}$ term describes the
hybridization between the upper Hubbard and Zhang-Rice singlets. The
last term results from the Coulomb repulsion between a Cu
$3d_{x^{2}-y^{2}}$ and its neighboring O $2p_{x,y}$ electrons.
$V_{pd}$ is proportional to the Coulomb repulsion between Cu and O
ions $U_{pd}$.

The above Hamiltonian can be simplified if $t_{ij}^{e} = t_{ij}^{h}
= t_{ij}$. In this case, by using the holon-doublon representation
of an electron operator $c_{i\sigma }= \sigma h_{i}^{\dagger }
d_{i\sigma } + e_{i} d_{i \overline{\sigma }}^{\dagger }$, and
taking a mean-field approximation for the $V_{pd}$-term,
$e_{i}^{\dagger }e_{i}h_{j}^{\dagger }h_{j} \approx \langle
e_{i}^{\dagger }e_{i} \rangle h_{j}^{\dagger }h_{j} + e_{i}^{\dagger
}e_{i} \langle h_{j}^{\dagger } h_{j} \rangle - \langle
e_{i}^{\dagger }e_{i} \rangle \langle h_{j}^{\dagger }h_{j} \rangle
$, one can then express $H$ as
\begin{equation}\label{eq:tUJ}
H=\sum_{ij\sigma }t_{ij}c_{i\sigma }^{\dagger }c_{j\sigma
}+U\sum_{i}n_{i\uparrow }n_{i\downarrow }+J\sum_{\langle ij\rangle
}S_{i}\cdot S_{j},
\end{equation}
where $n_{i\sigma }=c_{i\sigma }^{\dagger }c_{i\sigma }$ and
$U=\varepsilon_{e} + \varepsilon _{h}-4V_{pd}( \langle
e_{i}^{\dagger } e_{i}\rangle +\langle h_{i}^{\dagger } h_{i}
\rangle )$. In electron doped materials, as the induced hole
concentration is very small, $\langle h_{i}^{\dagger } h_{i}\rangle
\ll \langle e_{i}^{\dagger } e_{i}\rangle \approx x $, we have
$U\approx \varepsilon_{e} + \varepsilon _{h}-4 x V_{pd}$. It should
be emphasized that the spin exchange term in Eq. (\ref{eq:tUJ}) is
not a derivative of the one-band Hubbard model in the strong
coupling limit. It actually arises from the antiferromagnetic
superexchange interaction between two undoped Cu$^{2+}$ spins via an
O $2p$ orbital. This term, as shown in Ref. \cite{Daul00}, can
enhance strongly the superconducting pairing potential.

The $t$-$U$-$J$ model defined by Eq. (\ref{eq:tUJ}) is obtained by
assuming $t_{ij}^{e}=t_{ij}^{h}=t_{ij}$. This is a strong
approximation which may not be fully satisfied in real materials.
Nevertheless, we believe that this simplified model still catches
qualitatively the low energy physics of high-T$_c$ cuprates. It has
already been used, as an extension of either the Hubbard or the
$t$-$J$ model, to explore physical properties of strongly correlated
systems, such as the gossamer superconductivity.\cite{Zhang03}

The above Hamiltonian reveals two features about the effective
Hubbard interaction. First, $U$ is determined by the CT
gap\cite{Zaanen85}, not the Coulomb interaction between two
electrons in a Cu $3d_{x^2-y^2}$ orbital. It is in the intermediate
or even weak coupling regime, rather than the strong coupling limit
as usually believed. Second, $U$ is doping dependent. It drops with
doping. These are in fact the two key features that are needed in
order to explain the experimental results with the Hubbard model
\cite{Kusko02, Kusunose03, Senechal04}.

We have calculated the single-particle spectral function and the
staggered magnetization for the $t$-$U$-$J$ model using the
mean-field approximation. In the calculation, $t_{ij}$ are
parameterized by the first, second and third nearest neighbor
hopping integrals $(t,t^{\prime },t^{\prime \prime })$. The
parameters used are $t=0.326$ eV, $t^\prime = -0.25 t$, $t^{\prime
\prime} =0.15 t$, $J = 0.3 t$ eV, $\varepsilon_{e} + \varepsilon
_{h} = 4 t$ and $V_{pd} = 2.7 t$.

Fig. \ref{fig:band} shows the intensity plot of the spectral
function at the Fermi level. The doping evolution of the Fermi
surface agrees with the ARPES measurements.\cite{Armitage02} It is
also consistent with the mean-field calculation of the Hubbard model
by Kusko et al.,  while our calculation has the same shortcoming of
the mean field calculation in providing too large band
width.\cite{Kusko02} The difference is that, in our calculation, the
Hubbard interaction $U$ is not an adjustable parameter of doping. It
decreases almost linearly with doping. For the parameters given
above, $U \approx 4t - 10.8xt$.  Whereas in the calculation of Kusko
et al.\cite{Kusko02}, $U$ is determined by assuming the mean-field
energy gap to be equal to the experimentally observed value of the
``pseudogap".

\begin{figure}[ht]
\begin{center}
\includegraphics[width=0.8\columnwidth]{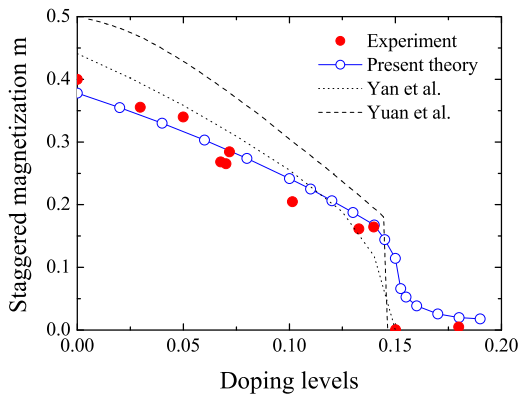}
\end{center}
\caption{Comparison of the mean field result (open circles) of the
staggered magnetization $m$ as a function of doping with the
experimental data (solid circles)\cite{stagger}. The theoretical
data obtained by Yan \emph{et al.}\cite{Yan06} and by Yuan \emph{et
al.}\cite{Yuan04} are also shown for comparison. }
\label{fig:stagger}
\end{figure}

Fig. \ref{fig:stagger} shows the theoretical result of the
staggered magnetization $m=(\langle n_{i\uparrow} -
n_{i\downarrow}\rangle )/2$. The simple mean-field result agrees
well with the experimental
data\cite{Mang04,Ross91,Matsuda90,stagger}, especially in the low
doping range. It is also consistent qualitatively with other
theoretical calculations\cite{Yan06,Yuan04}. $m$ decreases almost
linearly at low doping. However, it shows a fast drop above $\sim
0.14$, when the lower Zhang-Rice singlet holes begin to emerge
above the Fermi surface. This abrupt change of $m$ is an
indication of a significant renormalization of the Fermi surface.
It may result from the quantum critical fluctuation as suggested
in Ref. \cite{Dagan04}. $m$ does not vanish above the optimal
doping, this is probably due to the mean-field approximation.

In conclusion, we have shown that the minimal CT gap is much smaller
than the optical gap and the charge fluctuation between the
Zhang-Rice singlet and the upper Hubbard bands is strong in electron
doped copper oxides. The low-lying excitations of the system are
governed by the hybridized two-band $t$-$J$ model defined by Eq.
(\ref{eq:tj}) or approximately by the $t$-$U$-$J$ model defined by
Eq. (\ref{eq:tUJ}). This conclusion is drawn based on the analysis
of electron doped materials. However, we believe it can be also
applied to hole doped cuprate superconductors. Our mean-field
calculation for the $t$-$U$-$J$ model gives a good account for the
doping evolution of the Fermi surface as well as the staggered
magnetization. It sheds light on the further understanding of
high-T$_c$ superconductivity.

We wish to thank N. P. Armitage and  N. L. Wang for providing the
ARPES and infrared conductivity data shown in Fig.~\ref{fig:infra}.
Support from the NSFC and the national program for basic research of
China is acknowledged. The Stanford work was supported by DOE Office
of Science, Division of Materials Science, with contract
DE-AC02-76SF00515.


\begin{thebibliography}{9}
\bibitem{stagger} The staggered magnetization $m$ is determined by
multiplying the value of $m(x)/m(0)$ given in Ref. \cite{Mang04}
with the staggered magnetization at zero doping, $m(0)=0.4$, given
in Ref. \cite{Matsuda90}.
\end{thebibliography}
\end{document}